\documentclass{article}
\pdfoutput=1
\usepackage{arxiv}

\usepackage[utf8]{inputenc} 
\usepackage[T1]{fontenc}    
\usepackage{hyperref}       
\usepackage{url}            
\usepackage{booktabs}       
\usepackage{amsfonts}       
\usepackage{nicefrac}       
\usepackage{microtype}      
\usepackage{amsmath}
\usepackage{cleveref}       
\usepackage{lipsum}         
\usepackage{graphicx}
\usepackage[numbers]{natbib}
\usepackage{doi}

\usepackage{booktabs}
\usepackage{subcaption}
\usepackage{multirow}
\usepackage{enumitem}
\title{Modelling the Dynamics of Identity and Fairness in Ultimatum Game}

\date{}

\newif\ifuniqueAffiliation
\uniqueAffiliationtrue

\ifuniqueAffiliation 
\author{ Janvi Chhabra \\
	International Institute of Information Technology, Bangalore\\
	\texttt{janvi.chhabra@iiitb.ac.in} \\
	\And
	Jayati Deshmukh \\
	International Institute of Information Technology, Bangalore\\
	\texttt{jayati.deshmukh@iiitb.org} \\
 \And
    Arpitha Malvalli \\
	International Institute of Information Technology, Bangalore\\
	\texttt{arpitha.malavalli@iiitb.ac.in} \\
 \And
    Karthik Sama \\
	International Institute of Information Technology, Bangalore\\
	\texttt{sai.karthik@iiitb.ac.in} \\
 \And
	Srinath Srinivasa \\
	International Institute of Information Technology, Bangalore\\
	\texttt{sri@iiitb.ac.in} \\
}
\else
\usepackage{authblk}

\newbox{\orcid}\sbox{\orcid}{\includegraphics[scale=0.06]{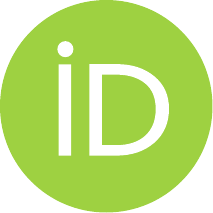}} 
\author[1]{%
	\href{https://orcid.org/0000-0000-0000-0000}{\usebox{\orcid}\hspace{1mm}David S.~Hippocampus\thanks{\texttt{hippo@cs.cranberry-lemon.edu}}}%
}
\author[1,2]{%
	\href{https://orcid.org/0000-0000-0000-0000}{\usebox{\orcid}\hspace{1mm}Elias D.~Striatum\thanks{\texttt{stariate@ee.mount-sheikh.edu}}}%
}
\affil[1]{Department of Computer Science, Cranberry-Lemon University, Pittsburgh, PA 15213}
\affil[2]{Department of Electrical Engineering, Mount-Sheikh University, Santa Narimana, Levand}
\fi

\hypersetup{
pdftitle={Modelling the Dynamics of Subjective Identity in Allocation Games},
pdfsubject={cs.MAS},
pdfauthor={Janvi Chhabra, Jayati Deshmukh, Srinath Srinivasa},
pdfkeywords={Identity, Allocation, Allocation game, Responsibility, Fairness},
}

\begin{document}
\maketitle
\begin{abstract}
  Allocation games are zero-sum games that model the distribution of resources among multiple agents. In this paper, we explore the interplay between an \textit{subjective identity} and its impact on notions of fairness in allocation. The sense of identity in agents is known to lead to responsible decision-making in non-cooperative, non-zero-sum games like Prisoners' Dilemma, and is a desirable feature to add into agent models. However, when it comes to allocation, the sense of identity can be shown to exacerbate inequities in allocation, giving no rational incentive for agents to act fairly towards one another. This lead us to introduce a sense of fairness as an innate characteristic of autonomous agency. For this, we implement the well-known Ultimatum Game between two agents, where their sense of identity association and their sense of fairness are both varied. We study the points at which agents find it no longer rational to identify with the other agent, and uphold their sense of fairness, and vice versa. Such a study also helps us discern the subtle difference between responsibility and fairness when it comes to autonomous agency. 
\end{abstract}


\section{Introduction}

The allocation of limited resources amongst individuals or groups with competing needs often creates a dilemma especially when conflicting goals, interests, or values are involved. This can be classified under a broad umbrella of responsibility dilemma, where an agent faces a conflict between an individually optimal state and a collectively or socially optimal state. These dilemmas are relevant in the context of both individuals in human societies as well as autonomous agents in multi-agent systems \cite{chevaleyre2005}.

In such dilemmas, the classical game theory models agents to act rationally i.e. they are modelled to choose actions which maximise their payoff. However, often human societies don't demonstrate such selfish behaviour; humans in such dilemmas do consider factors more than just their personal benefit~\cite{sen1977}. Responsible behaviour often emerges in human populations even in times of extreme conflict and oppression~\cite{bregman2020}. Such behaviour indicates that agents' decision-making should be more nuanced, maximising not just their own payoff but also being aware of the consequences of their actions on others.

The notion of fairness in agents has been widely studied in resource allocation scenarios~\cite{bin2020}. The notion of fairness of an agent attributes the allocation which an agent perceives as \textit{fair}. 

Humans facing the dilemma in allocation games demonstrate a preference towards fairness in addition to personal benefit~\cite{de2008}. They perceive the allocation through the lens of their notion of fairness and hence the utility function which they try to maximise is also influenced by their fairness preference. 

The notion of fairness in agents has been modelled in diverse ways. In some models, agents' notion of fairness is modelled as conforming with norms such as inequity aversion and reciprocity, where adhering to these norms is incorporated into the agents' utility function~\cite{fehr1999}. Another way the notion of fairness has been approached is by considering the influence of emotion and cognition in agents' utility functions~\cite{tamarit2016}. 
Reinforcement learning based approaches have also been used by agents to learn human strategies in the scenario of an ultimatum game. 

We extend a recently proposed model of responsible agency called Computational Transcendence (CT) which models an elastic identity in autonomous agents and it has been demonstrated to result in emergent responsible behaviour~\cite{deshmukh2022}. We introduce a notion of fairness into this model and then agents using this extended model are evaluated in the context of allocation games.

In this paper, we specifically look at the Ultimatum Game (UG) as an allocation scenario. This game highlights the different aspects involved in the decision-making process such as the notion of fairness in agents. The decisions of agents in social dilemmas such as UG have been interpreted in various ways, and in this work, we explore how responsible autonomous agents with a model of fairness act specifically in the context of the ultimatum game.

\section{Related Work}
In this section, we discuss in detail the ultimatum game (UG) and elaborate rational human behaviour in such a scenario. We also describe some existing approaches to model fairness in the ultimatum game. Further, we discuss a few frameworks that point to the importance of the subjective notion of fairness in the agents. Finally, we introduce the existing CT model and introduce fairness in this model in the context of UG.

\subsection{Ultimatum Game}

\begin{figure*}[!htb]
\centering
\includegraphics[width=0.7\linewidth]{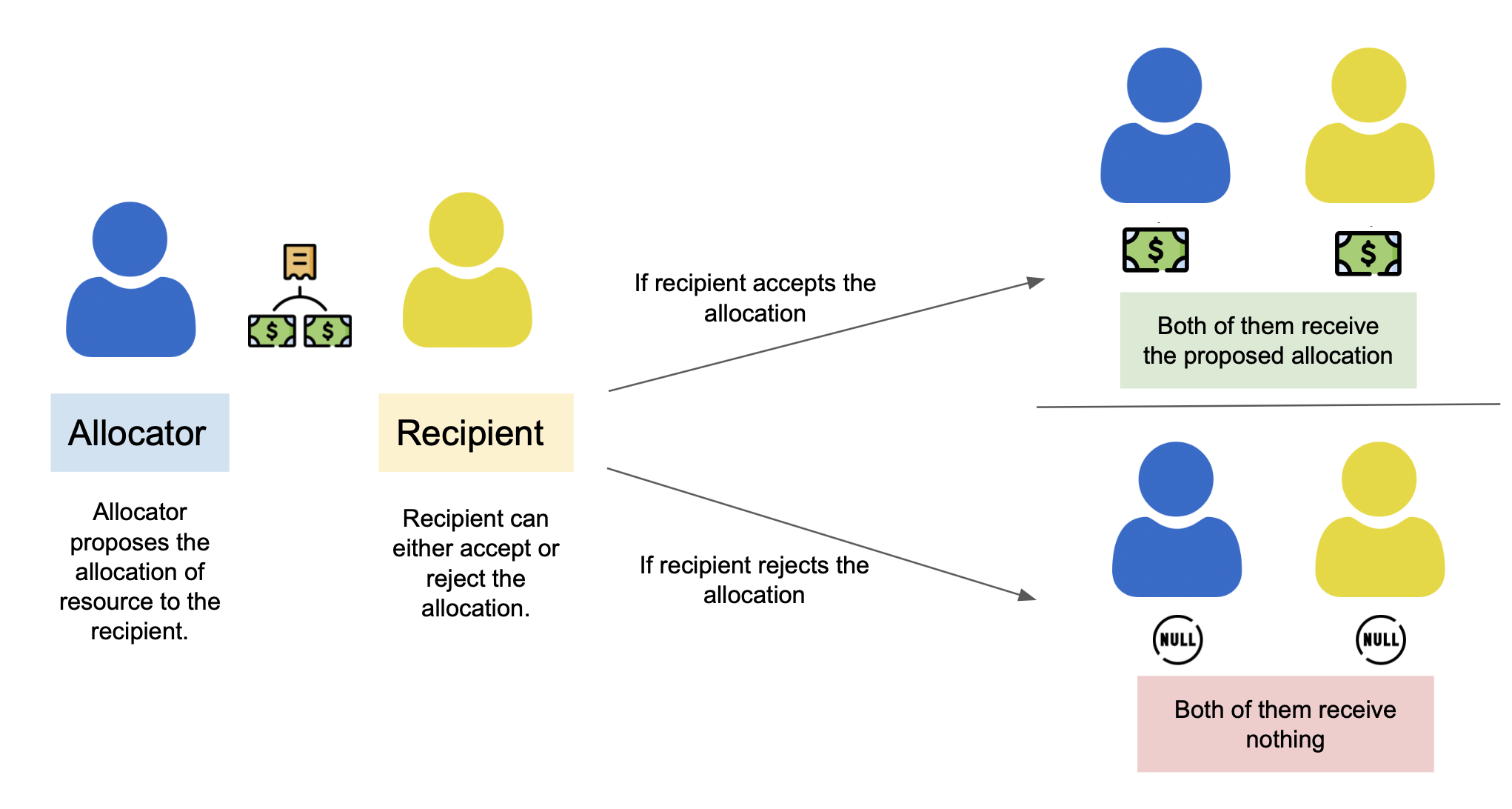}
\caption{The Ultimatum Game}
\label{fig:UG}
\end{figure*}

The Ultimatum Game (UG) was introduced by three economists Güth, Schmittberger, and Schwarze in 1982~\cite{guth1982}. In the standard UG between two agents, there exist two roles: Allocator and Recipient. For a limited resource $R$, the allocator proposes the allocation of the resource amongst itself and the recipient. Then the recipient can either accept or reject the proposed allocation.

\begin{itemize}[leftmargin=*]
    \item If the recipient \textbf{accepts} the proposed allocation, both allocator and recipient receive the split proposed by the allocator. 
    \item If the recipient \textbf{rejects} the proposed allocation, both allocator and recipient receive nothing.
\end{itemize}

Figure~\ref{fig:UG} summarises the ultimatum game between the allocator and the recipient. This game has led to significant research in analysing the trade-off between personal and comparative outcomes~\cite{larrick1997}. It has been introduced to test the theoretical assumption that utility maximisation can be equated to maximising personal monetary payoffs~\cite{guth1982}. Also, it has given rise to numerous other games like the dictator game, trust game, gift exchange game, and public goods game~\cite{van2014}.

\subsection{UG in Humans}
If the players are assumed to be rational as per the classical rational theory, it means that players will optimise to maximise their external payoff. Following this assumption, as per the game equilibrium, the allocator should offer the smallest possible split of the resource to the recipient, as it knows that the recipient can either choose between obtaining the proposed split by accepting or obtain nothing by rejecting the deal. And it makes rational sense for the recipient to accept the proposed split as accepting the deal would maximise its payoff. 

However, this rational assumption often is not observed in the case of experimental studies conducted on humans. The experimental results, when humans played the ultimatum game as a part of the experimental setup suggest that the modal offer that the allocator proposes is often around 50$\%$ and offers $\leq$ 20$\%$ have around 50$\%$ chance of getting rejected~\cite{camerer2011} by the recipient. These findings suggest that humans don't just maximise their external payoff, but are also concerned about the comparative payoff, which indicates that they have a notion of fairness. UG became a popular game in experimental studies to understand the behaviour of allocator and recipient when humans play these games. A few of the scenarios which were experimentally tested are as follows:

\begin{itemize}[leftmargin=*]
    \item \textbf{When the stake of the resource to be allocated is increased}: Even a 20 times increase in stake didn't have a significant impact on allocator and recipient behavior~\cite{list2000}. 
    \item \textbf{When the players were known or anonymous to each other}: When the players were known to each other, the allocator usually offered a greater split to the recipient as compared to when they were anonymous~\cite{bolton1995}.
    \item \textbf{When UG was played across various cultures}: It was observed that variation in culture leads to very diverse behaviour. In most dominant cultures, the game settles in the state as discussed so far. However, in some cultures, the settled state of the game is different from usual. For example, in the Peruvian tribe in the Amazon, the allocator offers a very low share (around 26$\%$) to the recipient and faces a very high rejection rate from recipients~\cite{roth1991}. In contrast to this, in the case of Lamelara tribes of Indonesia, the allocator offers a very high share (around 58$\%$) to the recipient and yet approximately 37$\%$ of such deals are rejected by recipients~\cite{cameron1999}. 
\end{itemize}

This evidence from experimental studies suggests that humans in the UG scenario do not behave purely based on the rational (payoff maximization) assumption. The evidence, especially from cultural studies indicates that the comparative payoff is interpreted in diverse ways across cultures. This also indicates the effect of culture on the identity of humans, which influences how they perceive such UG scenarios. 

\subsection{Modelling Fairness in UG}

Next, we discuss various ways in which the modelling of agents' behaviour has been approached specifically in the UG scenario.

\subsubsection{Inequity Aversion based Model} \label{subsec:inequity-averse}
Fehr and Schmidt~\cite{fehr1999} proposed an inequity-aversion-based model of fairness in the UG context. An agent is inequity-averse when it dislikes the inequitable allocation of resources irrespective of which side of the inequity it lies on. This inequity-averse nature of agents affects their utility function. They assume that being on the disadvantageous side of inequity gives a more negative utility as compared to being on the advantageous side of the inequity. 

\subsubsection{Reciprocity based Model}
Rabin, Falk and Fischbacher proposed a reciprocity-based model of fairness~\cite{rabin1993,falk2006}. Using this model, agents evaluate the ``kindness'' of the action of the other agent they are interacting with and respond or reciprocate accordingly. This evaluation of kindness constitutes factoring in the consequences of the other agent's actions and the underlying intentions behind those actions. 

The utility function of the agents is influenced by the kindness of the action and the reciprocation reaction of agents to that treatment. Agents' actions are influenced by their first-order and second-order beliefs. They try to maximise their utility and align their actions with their beliefs. 

\subsubsection{Emotion and Cognition based Model}
Tamarit and Sánchez~\cite{tamarit2016} proposed a model where agents' behaviour is influenced by a combination of both their emotions and cognition. It is inspired by Kanheman's two-system theory~\cite{kahneman2002}, which proposes that the decision-making process is a result of the interaction between two systems; System 1 which is fast, intuitive, associative, and emotionally driven and System 2 which is slow, cognitively demanding, and analytical. 

\subsubsection{Social Decision-Making in Ultimatum Game}
The Social Utility framework proposed for UG~\cite{handgraaf2003} categorises the factors which influence the behaviour of agents in the game. These broad categorisations are as follows:

\begin{itemize}[leftmargin=*]
    \item \textbf{Contextual Factors}: Contextual factors in UG are changing the nature of the resource to be allocated, framing the rules of the game, and the game being played across diverse sets of people.
    \item \textbf{Characteristics of Players}: This can be attributed to the social distance between the players i.e. whether they are known or anonymous to each other. This can also account for the power dynamics between the players and the personality of the players which can influence the motivation of the players.  
    \item \textbf{Characteristics of the Game}: This is attributed to the way in which the game is structured. Different variants of UG, such as the dictator game can also give an idea about the factors of the game itself which impact certain behaviour of players. 
\end{itemize}

\subsection{Computational Transcendence}\label{sec:CT}

Computational Transcendence (CT)~\cite{deshmukh2022} models an elastic identity or a sense of self in autonomous agents using which they can identify with external entities like other agents, groups and notions in the system. The sense of self of an agent $a$ is represented as $S(a) = (I_a, d_a, \gamma_a)$ where $I_a$ represents the identity set of the agent consisting of aspects it identifies with, $d_a$ is the semantic distance of the agent which denotes the perceived logical distance of an agent to each aspect in its identity set and $\gamma_a$ is the transcendence level of the agent which denotes the extent to which it identifies with others. An agent $a$, with transcendence level $\gamma_a$ identifies with an aspect $o$ whose distance is $d_a(o)$ with an attenuation factor of $\gamma_a^{d_a(o)}$. It has been shown that this model leads to emergent responsible behaviour by autonomous agents across diverse scenarios. In this paper, first, we test transcended agents in the UG setup and then extend the model by incorporating a notion of fairness in transcended agents.

Identifying with external entities affects how an agent's internal valuation or \textit{utility} is computed based on external rewards or \textit{payoffs} that may be received by different aspects in its identity set. For any aspect $o \in I_a$, let the term $\pi_i(o)$ refer to the payoff obtained by aspect $o$ in the game or system state $i$. Given this, the utility derived by agent $a$ in system state $i$ is computed as follows:

\begin{equation}
    \begin{split}
    u_i(a) & =  \frac{1}{\sum_{\forall o \in I} \gamma_a^{d_a(o)}} \displaystyle \sum_{\forall o \in I} \gamma_a^{d_a(o)} \pi_i(o)
    \end{split}
   \label{eqn:ctutility}
\end{equation}

\section{Baseline Model} \label{subsec:ug}
Consider the scenario, where the resource $R$ is assigned the value $1$, and it is to be divided among two agents with the roles of \textit{Allocator} and \textit{Recipient}. The allocator proposes the allocation of the resource valued $1$ as $x$ for itself and $1-x$ for the recipient. Next, the recipient can either accept or reject the proposed allocation. If the recipient accepts this allocation, both allocator and recipient receive the resource (referred to as payoff) of $x$ and $1-x$ respectively. On the other hand, if the recipient rejects the proposed allocation, then both of them get nothing i.e. the payoff for both agents is $0$. 

As discussed, the rational strategy using payoff maximization for the allocator is to propose a minimum possible split of resource $R$ to the recipient, and the recipient accepts it because rejecting the allocation results in a payoff of $0$.

Transcended agents make the decision based on the expected utility of their choices. Suppose the two transcended agents playing UG are represented as follows: a transcended agent $a_{1}$ having a transcendence level of $\gamma$ and semantic distance $d$ with the other agent $a_{2}$. The utility of $a_{1}$, when it receives the split of $x$ and $a_{2}$ receives the split of $1-x$, is computed in Equation~\ref{eq:oldTrans} as follows:

\begin{equation}
    util(a_{1}) = \frac{x + \gamma^{d}(1-x)}{1 + \gamma^{d}}
    \label{eq:oldTrans}
\end{equation}

\noindent In this scenario,
\begin{itemize}[leftmargin=*]
    \item \textbf{If $a_{1}$ is the allocator}: It proposes the maximum utility split, i.e. it computes the utility for all the possible splits and proposes the split that gives it the maximum utility. 
    \item \textbf{If $a_{1}$ is the recipient}: It accepts if the proposed split gives $\geq$ its minimum acceptable utility i.e. it calculates the utility of the proposed split and if it is $\geq$ 0 it accepts, else it rejects the proposal. 
\end{itemize}

\begin{figure}
\centering
\includegraphics[width=0.6\linewidth]{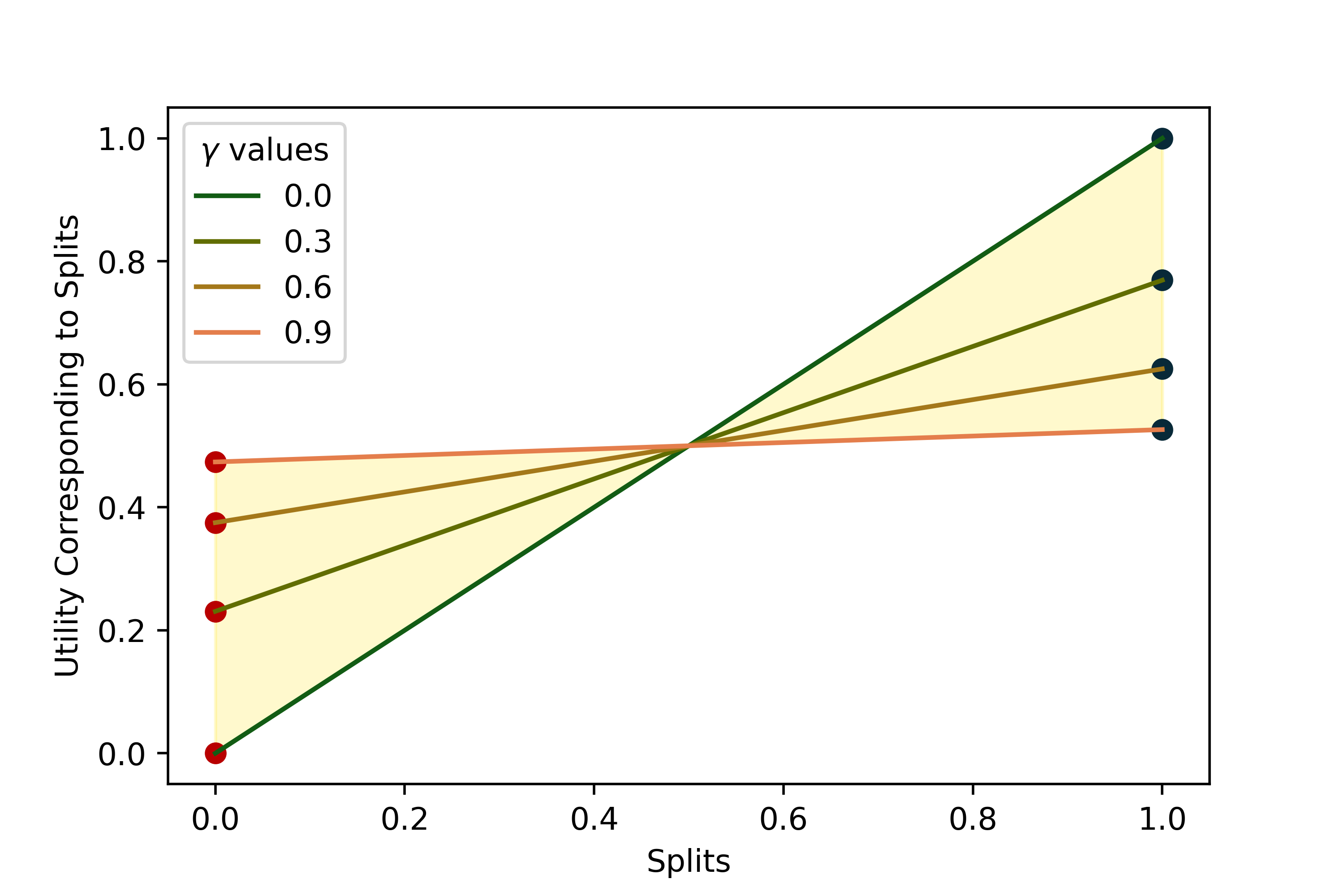}
\caption{Utility of a transcended agent for possible splits for various $\gamma$ values. The blue circles are Maximum Utility Split and the red circles are Minimum Acceptable Utility}
\label{fig:oldTrans}
\end{figure}

Figure~\ref{fig:oldTrans}, shows the utility of the transcended agent for all possible splits when the utility is computed using Equation~\ref{eq:oldTrans} for various transcendence ($\gamma$) values. We can infer from the plots that the transcended agent as an allocator takes the whole resource for itself (except at $\gamma=1$) as it has maximum utility when it allocates the whole resource to itself, as denoted by the blue circles. Also, as a recipient, the utility accrued is $\geq$ 0, as observed by red circles in Figure~\ref{fig:oldTrans}, thus it makes sense to always accept the proposed allocation. Hence, the game state for almost all values of $\gamma$ values involves the allocator taking almost the whole resource (in this case, the split of 1) and the recipient accepting the proposed allocation. According to the definition of transcendence, since the agents identify with each other, they also account for each other's payoff. However, in the absence of a notion of fairness, they propose and accept unfair allocations. Next, we introduce a notion of fairness in transcended agents and evaluate whether it leads to more diverse and fair allocations.

\section{Notion of Fairness}

The baseline model of transcendence does not result in fair and diverse deals in the case of allocation games. The more the agents transcend, the more they identify with the other agent, they find unfair deals acceptable since the other agent's utility also contributes to their utility. So we extend this model by introducing a notion of fairness in agents using a fairness threshold.

Fairness threshold, ($\tau$) is defined as a threshold below which if an agent receives an allocation, it perceives it to be unfair. The rationale for agents having a fairness threshold has been explored in the literature of UG, where agents reject any offer below their fairness threshold~\cite{henrich2000,oosterbeek2004,chang2012}. The fairness threshold acts as a lens through which the agents perceive the payoff they receive. The utility computation should account for the perceived payoff based on the fairness threshold instead of just the payoff received. Thus, the process of utility computation consists of two stages, first calculating the perceived payoff for the allocation and then computing the utility based on the perceived payoff with respect to the fairness threshold.

\subsection{Perceived Payoff}

\begin{figure}
\centering
\includegraphics[width=0.6\linewidth]{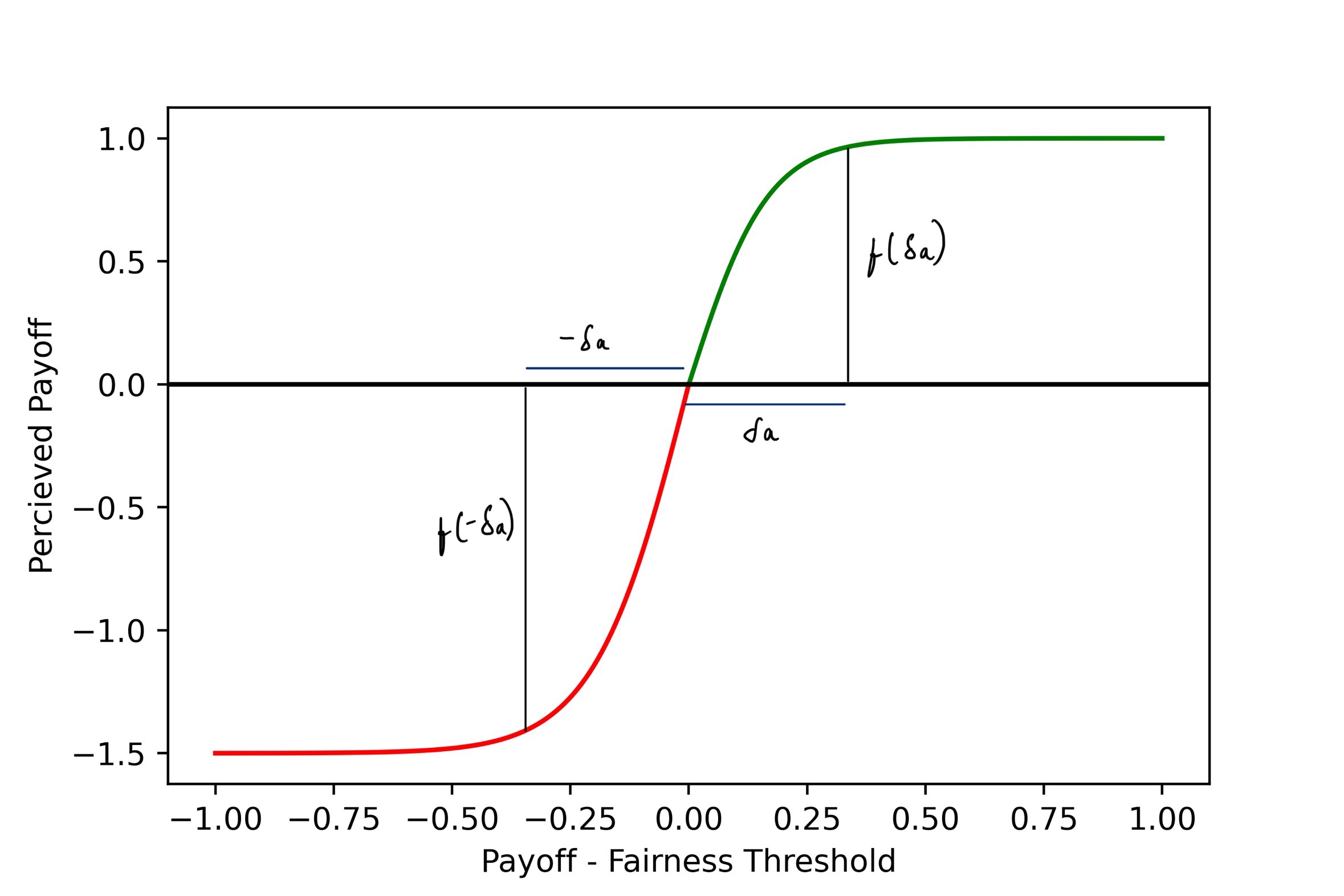}
\caption{Perceived Payoff Function}
\label{fig:ppFunc}
\end{figure}

The risk or loss aversion behaviour~\cite{kahneman2013} is usually demonstrated by humans. For the same value of loss or gain, humans perceive the absolute value of loss to be greater than the equivalent gain. Risk aversion is relevant, especially for agents having a notion of fairness and it can be used to compute the perceived payoff. Every agent has a fairness threshold ($\tau$) and it perceives the payoff it receives through the lens of its fairness threshold. An agent $a$ having a fairness threshold of $\tau_{a}$ receives allocation $x$. Its perceived payoff is computed using a sigmoid function as shown in Figure~\ref{fig:ppFunc}. The function $f(x-\tau_{a})$, which is a S-shaped function denotes the loss aversion behaviour of the agents~\cite{dacey2003}. The disparity between the received payoff, $x$ and fairness threshold $\tau_{a}$, is $\delta_{a}$. Then according to the function as shown in Figure~\ref{fig:ppFunc}, we can state that the inequality $\lvert f(\delta_{a}) \rvert < \lvert f(-\delta_{a}) \rvert$ holds. 

\subsection{Utility Computation}
As discussed, the external payoff agents receive is perceived through a lens of fairness threshold. Further, the agents compute their utility based on the perceived payoffs. As the utility computation also involves the payoff of the aspects the agent identifies with, it is also replaced by the perceived payoff. Hence, the agent not only perceives its payoff through the lens of its fairness threshold but also the split of the other agent it identifies with. 

Agents perceive the payoff through the lens of their $\tau$, thus this carries an assumption that other agents' fairness threshold is also the same as their fairness threshold. Hence, utility computation is a combination of perceived payoff (Figure~\ref{fig:ppFunc}) and transcended utility computation (Equation~\ref{eqn:ctutility}). Assuming that the total resource to be allocated is 1, the utility of the agent $a$, receiving the split $s$ is described in Equation~\ref{eq:combUtil} as follows:

\begin{equation}
    util(a) = \frac{f(s - \tau) + \gamma^d*f((1-s)-\tau)}{1 + \gamma^d} 
    \label{eq:combUtil}
\end{equation}

So far, the utility computation of the agents with a fairness threshold of $\tau$ was discussed. Next, we explore different possible ways to represent $\tau$ in agents. The following two alternatives for representing the fairness threshold in agents will be elaborated and experimental results for both will be discussed in detail:

\begin{itemize}
    \item \textbf{$\tau$ as an agent-based characteristic}: In this case, $\tau$ is a characteristic of the agent itself and it doesn't vary for different aspects in the identity set of the agent (similar to transcendence level, $\gamma$). Thus, no matter how close (low semantic distance) or far (high semantic distance) the agent is to some aspect in its identity set, its $\tau$ is the same. This model assumes that the fairness criteria of an agent is universal and same for everyone and thus it does not vary depending on whom the agent is interacting with. 
    
    \item \textbf{$\tau$ as an association-based characteristic}: In this case, the fairness threshold $\tau$ of an agent corresponds to each aspect in its identity set and it depends on how the agent associates with that aspect (similar to the semantic distance, d). In this case, there is a correlation between the semantic distance and fairness threshold for every aspect. This model assumes that fairness is not a universal characteristic but rather a notion that varies on a case-by-case basis depending on the interaction and perception of individual aspects in the identity set.
\end{itemize}

\section{Agent-based Fairness Threshold} \label{sec:staticFT}

In this section, we discuss in detail the decisions that the allocator and recipient make when they have fixed fairness threshold $\tau$. We also discuss the effect of varying other agent-specific parameters like $\tau$, $\gamma$, and $d$. As discussed in Section~\ref{subsec:ug}, the allocator proposes a split which gives it maximum utility and the recipient accepts a split if its utility is $\geq$ its minimum acceptable utility.

\subsection{Utility Plots}

\begin{figure}
\centering
\includegraphics[width=0.6\linewidth]{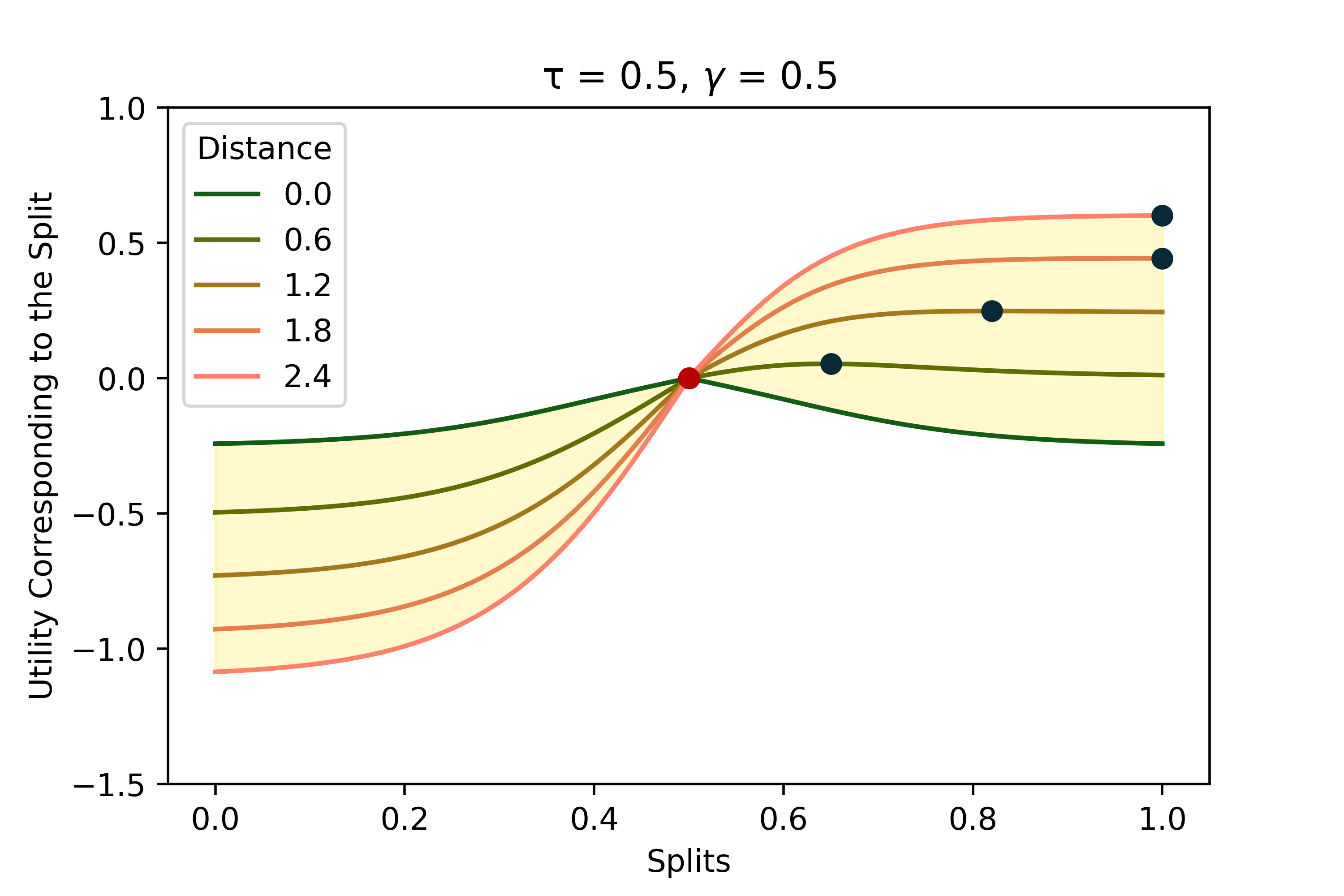}
\caption{The utility curve of the transcended agents ($\gamma$ = 0.5, $\tau$ = 0.5) w.r.t possible splits while varying the semantic distance of the agent with another player}
\label{fig:utilCurveHalf}
\end{figure}

Figure~\ref{fig:utilCurveHalf} shows the variation of the utility of a split for agents with $\gamma =0.5$ and $\tau = 0.5$. These utility curves are plotted for a range of semantic distances, $d$ of the agent with the other agent with whom it is playing the UG. The yellow-shaded region in the figure shows the spectrum of utility for possible splits when $d$ is varied between $0$ to $2.4$. 

The red circle shows the splits at which the recipient gets the minimum acceptable utility and the blue circles show the splits at which the allocator gets the maximum utility. From this plot, it can be inferred that with the increase in semantic distance $d$, the split which gives the maximum utility to the allocator also increases. Hence, as the allocator transcends higher, its semantic distances reduce and it is more likely to share the resource to a greater extent with the recipient. On the other hand, as the transcendence level decreases, the allocator is likely to take the whole resource for itself.

As the recipient, it is observed that varying the semantic distance does not have an impact on the acceptable split, as for all values of $d$, the split that satisfies the minimum acceptable utility is $0.5$. An interesting observation can be drawn from the nature of the curves with varying semantic distance. The curve for utility when $d=0$ is symmetric at the split of $0.5$, and as the distance increases it becomes S-shaped and gets positive utility from the being on the advantageous side of the split. The utility curve for $d=0$ shows the characteristic of an inequity-averse agent discussed earlier as it accrues negative utility being on either side of inequitable outcomes. As the value of $d$ increases, this nature of inequity-aversion decreases. 

Next, to understand the impact of $\tau$ and $\gamma$, we plot similar utility curves for different combinations of $\tau$ and $\gamma$, as shown in Figure~\ref{fig:utilCurves}. The effect of varying $\tau$ and $\gamma$ on the utility of the agent can be inferred as follows:

\begin{figure*}
    \centering
    \begin{subfigure}{0.45\textwidth}
         \centering
         \includegraphics[width=\textwidth]{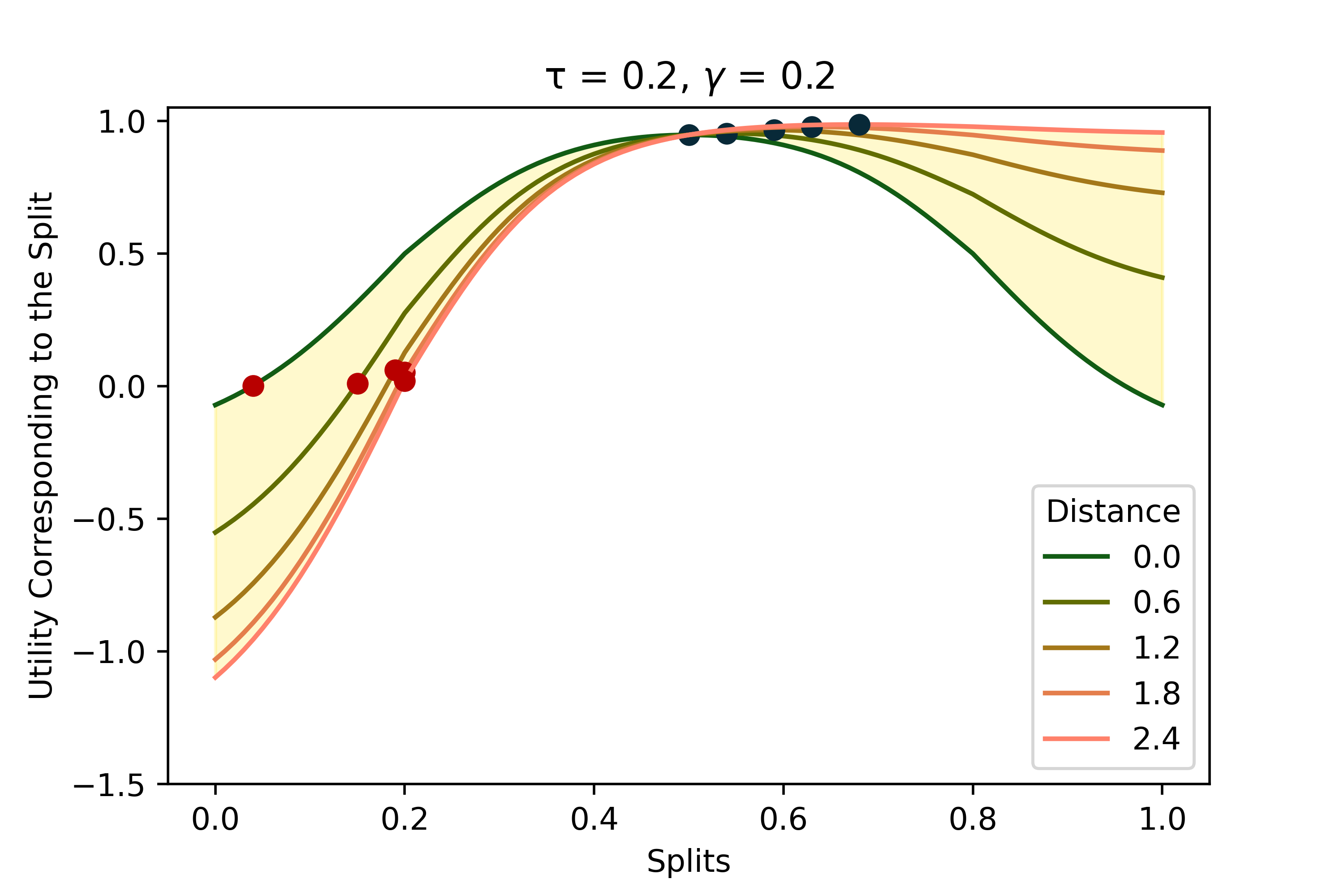}
    \caption{Low $\tau$ and Low $\gamma$}
    \label{fig:util-ltlg}
     \end{subfigure}
    \begin{subfigure}{0.45\textwidth}
         \centering
         \includegraphics[width=\textwidth]{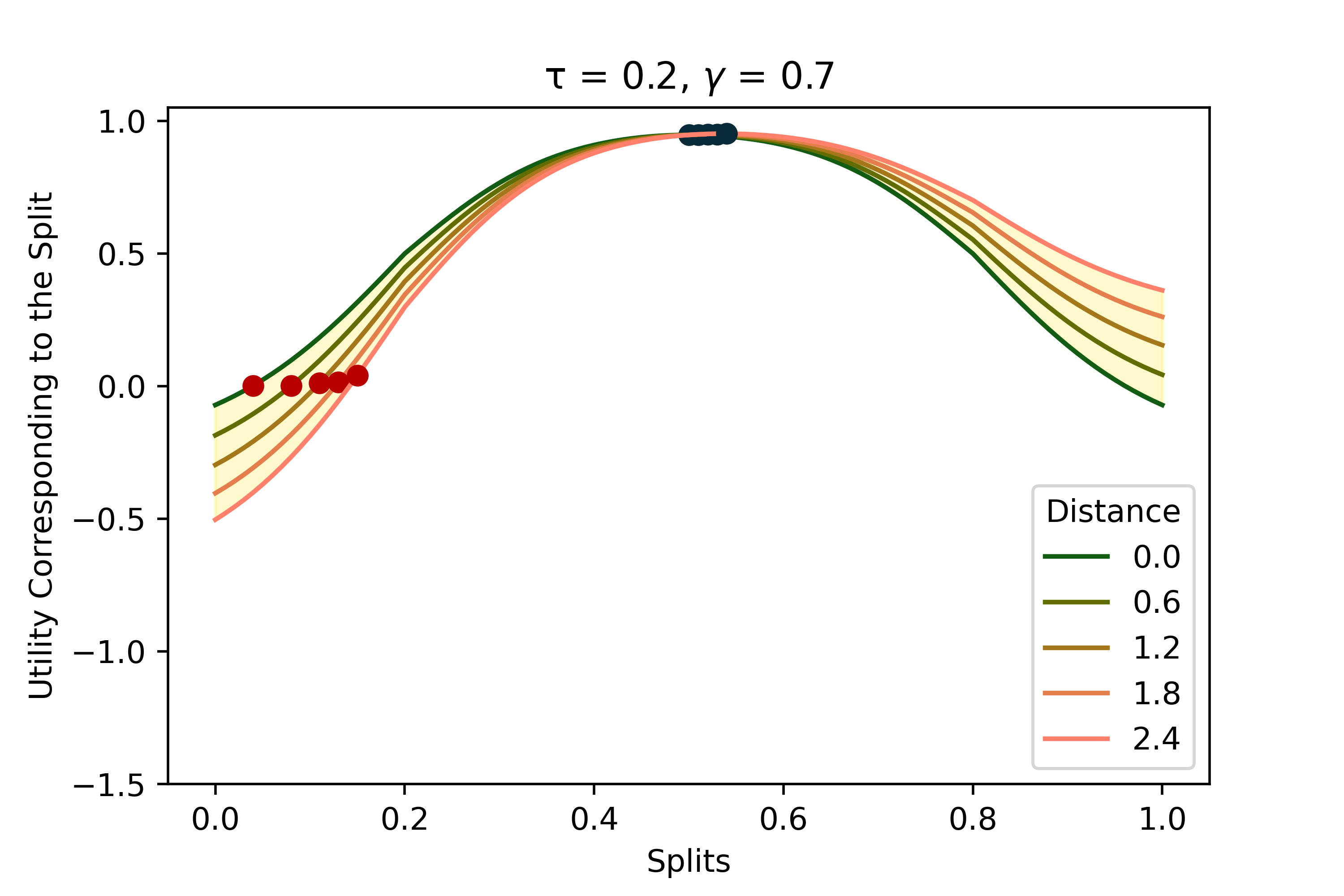}
    \caption{Low $\tau$ and High $\gamma$}
    \label{fig:util-lthg}
     \end{subfigure}
     
    \centering
    \begin{subfigure}{0.45\textwidth}
         \centering
         \includegraphics[width=\textwidth]{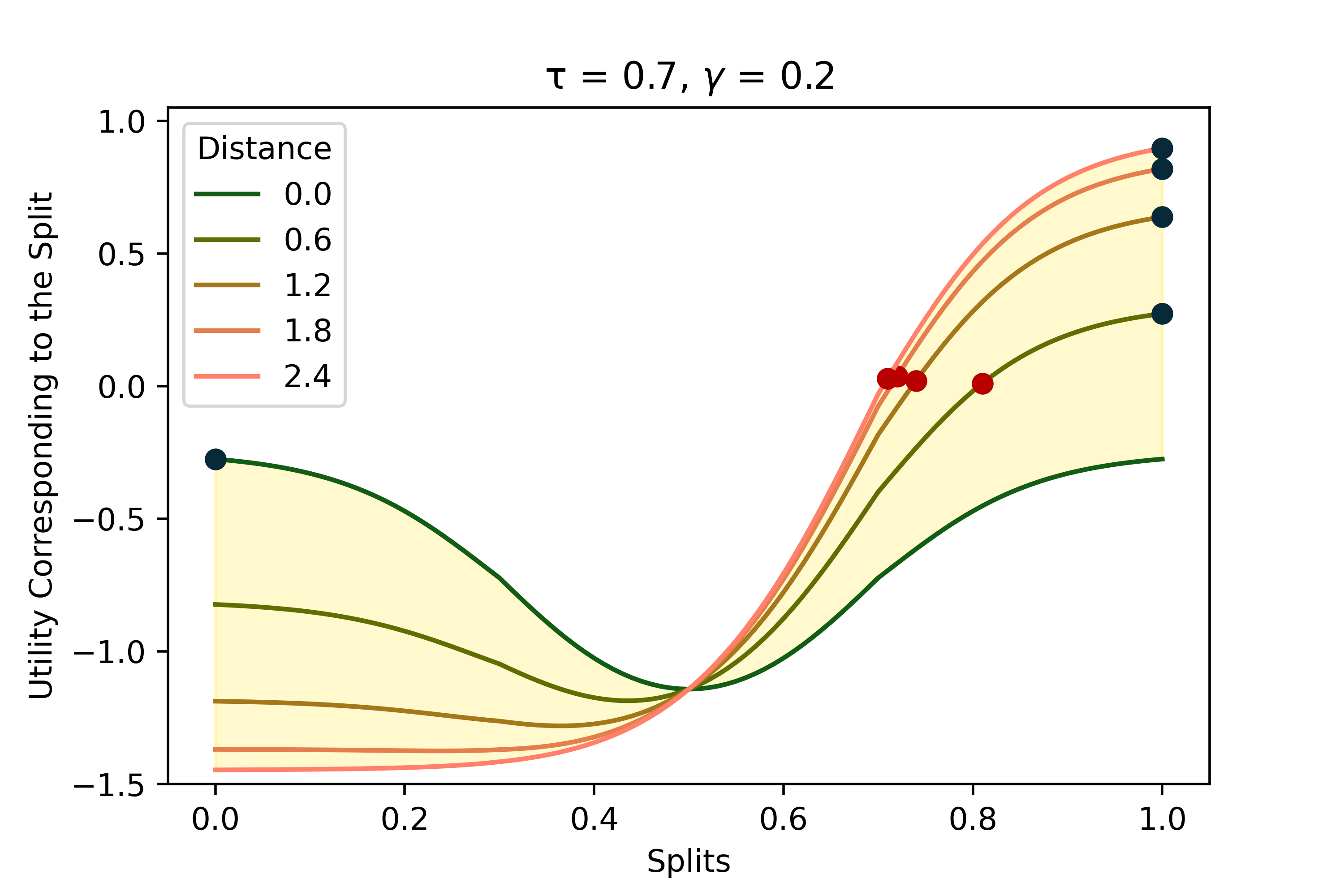}
    \caption{High $\tau$ and Low $\gamma$}
    \label{fig:util-htlg}
     \end{subfigure}
    \begin{subfigure}{0.45\textwidth}
         \centering
         \includegraphics[width=\textwidth]{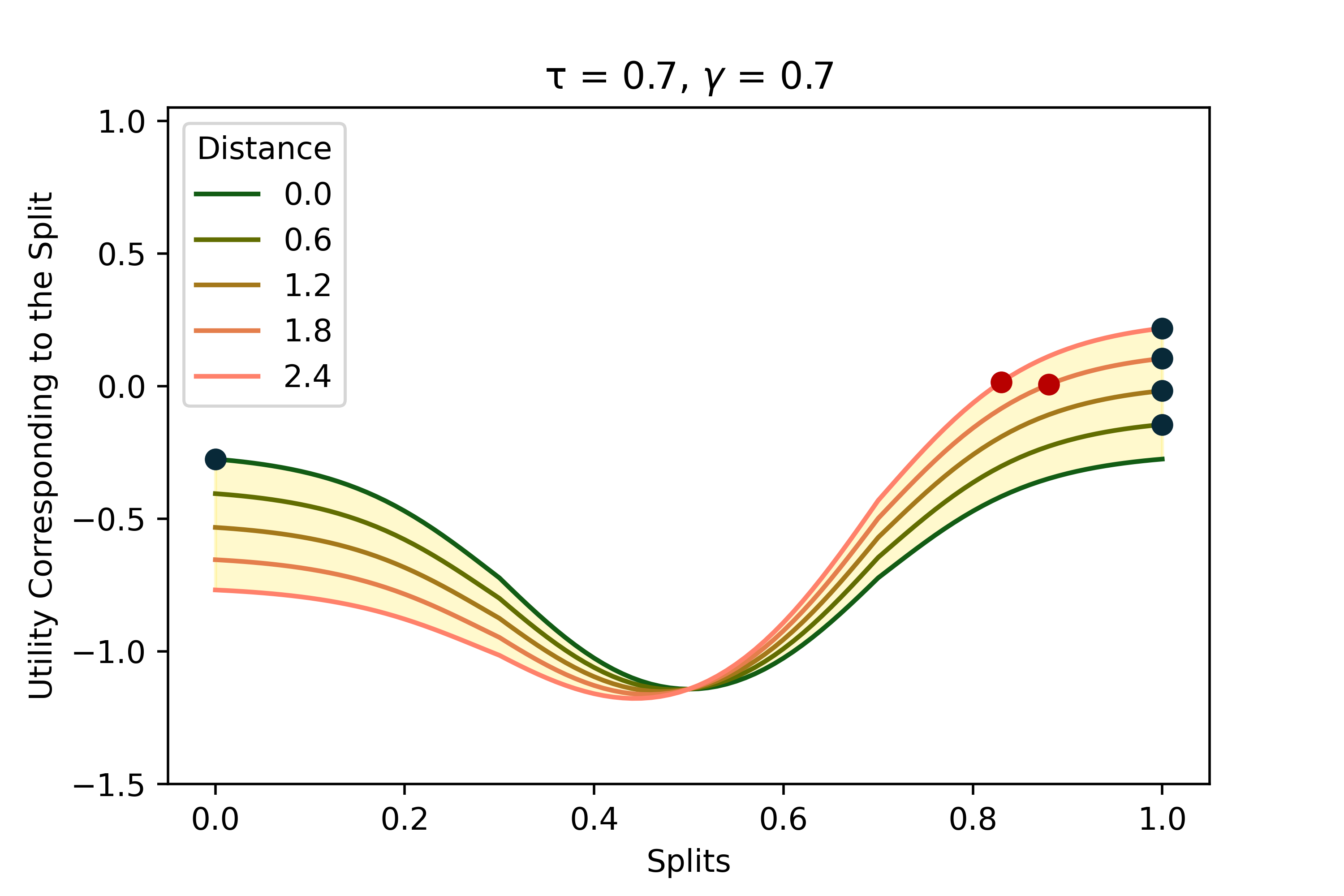}
    \caption{High $\tau$ and High $\gamma$}
    \label{fig:util-hthg}
     \end{subfigure}
    \caption{Utility curves with different combination for extremes values of $\tau$ and $\gamma$}
    \label{fig:utilCurves}
\end{figure*}

\begin{itemize}[leftmargin=*]
    \item \textbf{Effect of $\tau$ on utility}: \\ The nature of utility curves when $\tau$ is low is presented in Figure~\ref{fig:util-ltlg} and Figure~\ref{fig:util-lthg}. We note that the utility increases till the split of 0.5 and then it decreases (when $d$ is low), this is because the $\tau$ is 0.2, and the utility is maximum at 0.5 split of resource because both the agents are benefiting equally. For any split beyond 0.5, the other agent suffers more, and as observed for low $d$, the agent shows inequity-averse behaviour, hence, the utility of the agent decreases. This inequity-averse nature changes as the semantic distance is increased. 
    
    Figure~\ref{fig:util-htlg} and Figure~\ref{fig:util-hthg} show the utility plots when the value of $\tau$ is high. In the case of low $d$, the utility decreases as the split is increased till 0.5 split, where the agent has minimum utility and then it increases. This is because both agents are equally suffering in case of a 0.5 split and on either side one of them is benefiting. As the value of $d$ is increased, the agent transcends less, and the utility of the agent increases as the split is increased.

    In the case of low $\tau$, the allocator (represented by the blue circles) takes splits for itself which are near 0.5 and they increase the split as the distance with the recipient increases. In the case of the recipient (represented by the red circles), when the distance is less, they accept a split of low value, but it increases as the distance increases. Hence, in this case, many game states exist where the deal between the allocator and recipient is accepted.
    
    In the case of high $\tau$, an allocator is always motivated to take the whole split for itself (except $d=0$), (represented by the blue circles). And as a recipient, the split which gives the minimum acceptable utility is high around $0.7$ (represented by the red circles). This indicates that in this case, most of the deals are rejected (except when $d$ of allocator or recipient is 0 from the other player).

    \item \textbf{Effect of $\gamma$ on utility}: \\ When the $\gamma$ of the agent is low, as observed from Figure~\ref{fig:util-ltlg} and Figure~\ref{fig:util-htlg}, the splits which the allocator takes for itself (represented by the blue circles) increases rapidly with the increase in distance. Similarly, the splits at which recipients get the minimum acceptable utility (represented by the red circles) also increase rapidly compared to an agent with a high value of $\gamma$. 
    
    The effect of $\gamma$ can also be observed in terms of the extent of the yellow-shaded region. Agents with low $\gamma$ had more area of shaded region as compared to high $\gamma$ when the value of $\tau$ is kept constant. This shows that for the agents with low $\gamma$, the nature of the utility curve changes rapidly with varying $d$ as compared to agents with high $\gamma$ as shown in Figure~\ref{fig:util-lthg} and Figure~\ref{fig:util-hthg}. 
\end{itemize}

\subsection{Acceptable Splits for the Recipient}

\begin{figure*}
    \centering
    \begin{subfigure}{0.32\textwidth}
         \centering
         \includegraphics[width=\textwidth]{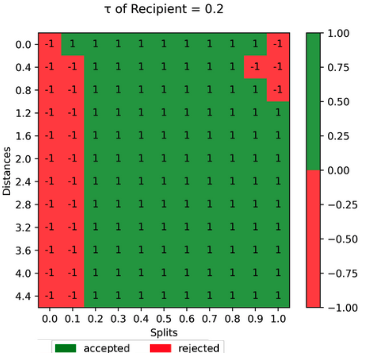}
    \caption{Low $\tau$}
    \label{fig:matStaticLowFT}
     \end{subfigure}
    \begin{subfigure}{0.32\textwidth}
         \centering
         \includegraphics[width=\textwidth]{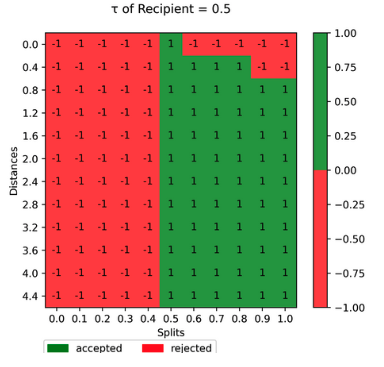}
    \caption{Moderate $\tau$}
    \label{fig:matStaticModFT}
     \end{subfigure}
     \begin{subfigure}{0.32\textwidth}
         \centering
         \includegraphics[width=\textwidth]{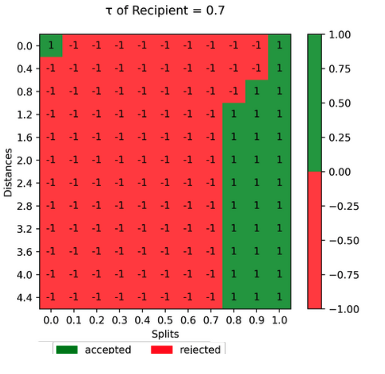}
    \caption{High $\tau$}
    \label{fig:matStaticHighFT}
     \end{subfigure}
     \caption{Acceptable Split Matrix with varying semantic distance of recipient with allocator (recipient with $\gamma=0.4$)}
    \label{fig:matRecStatic}
\end{figure*}

So far, we discussed the utility plots and the variation in the decisions of the allocator and recipient with the change in $\gamma$, $d$, and $\tau$. Next, we evaluate the game setup specifically focusing on the conditions which lead to deals being accepted. We analyse the game from the recipient's perspective and study the factors which affect the acceptance or rejection of the proposed deal. As noted earlier, $\gamma$ does not have a significant impact on the nature of the utility curve, so we choose a fixed value of $\gamma$ ($0.4$ in this case). We vary the semantic distance $d$ of the recipient from the allocator and analyse its impact on the acceptance or rejection of the split offered to the recipient. 

The matrices in Figure~\ref{fig:matRecStatic} represent the split on the x-axis and the variation in the semantic distance of the recipient with the allocator on the y-axis, for different values of $\tau$. The acceptance of a split is represented by green colour and rejected splits by red colour. The following inferences can be drawn from the matrices:

\begin{itemize}
    \item \textbf{Recipient with low $\tau$}: In Figure~\ref{fig:matStaticLowFT}, when the $\tau$ of the agent is low ($\tau$=0.2), it accepts all the deals where the recipient and allocator are getting  $\geq$ 0.2 up to a certain distance. After that recipient starts accepting deals in which the allocator is getting less than 0.2.

    \item \textbf{Recipient with moderate $\tau$}: For the agents with moderate $\tau$ ($\tau$=0.5), as shown in Figure~\ref{fig:matStaticModFT}, when the semantic distance of the recipient with the allocator is very low, it only accepts the 0.5 splits, as either way getting more or less than that is unfair for one of them. As the semantic distance with the allocator increases, it starts accepting all the splits more than 0.5.

    \item \textbf{Recipient with high $\tau$}: As observed in the above cases, recipients reject the higher split offers when the semantic distance with the allocator is less. This is because they also account for the fairness of the allocation of the allocator. Hence, when the agent's $\tau$ is high ($\tau$ = 0.7), and the semantic distance is more, they accept the extreme splits, where at least one of them is satisfied. As the distance increases, they accept the splits which are strictly greater than 0.7. 
\end{itemize}

In this section, the fairness threshold ($\tau$) is modelled as a fixed characteristic of the agent. The agents perceive the allocations through the lens of this threshold (factoring in their share as well as other player's share). As observed in Figure~\ref{fig:matStaticHighFT}, most of the deals are rejected when the agent's $\tau$ is high, even though the extent to which the agent transcends ($\gamma^{d}$) is high. Hence, in this model, we observe a contradiction in having a high fairness threshold as the innate characteristic and also highly transcending with the other agent. For example, a realistic scenario of resource allocation with high transcendence is sharing food with a family member. With a higher transcendence level, there is a higher tendency to share since the agent prefers to a higher extent the well-being of others. In this case, it doesn't make sense to have a fairness threshold of $0.7$ (i.e. taking atleast 70$\%$ of the food for oneself). In this example, having a static $\tau$ makes sense only in exceptional cases like when health requirements mandate having a minimum of 70$\%$ of the food. Next, we explore an alternate formulation where the fairness threshold ($\tau$) of an agent depends on the association of the agent with the other agent it interacts with.

\section{Association-based Fairness Threshold}

\begin{figure}
\centering
\includegraphics[width=0.5\linewidth]{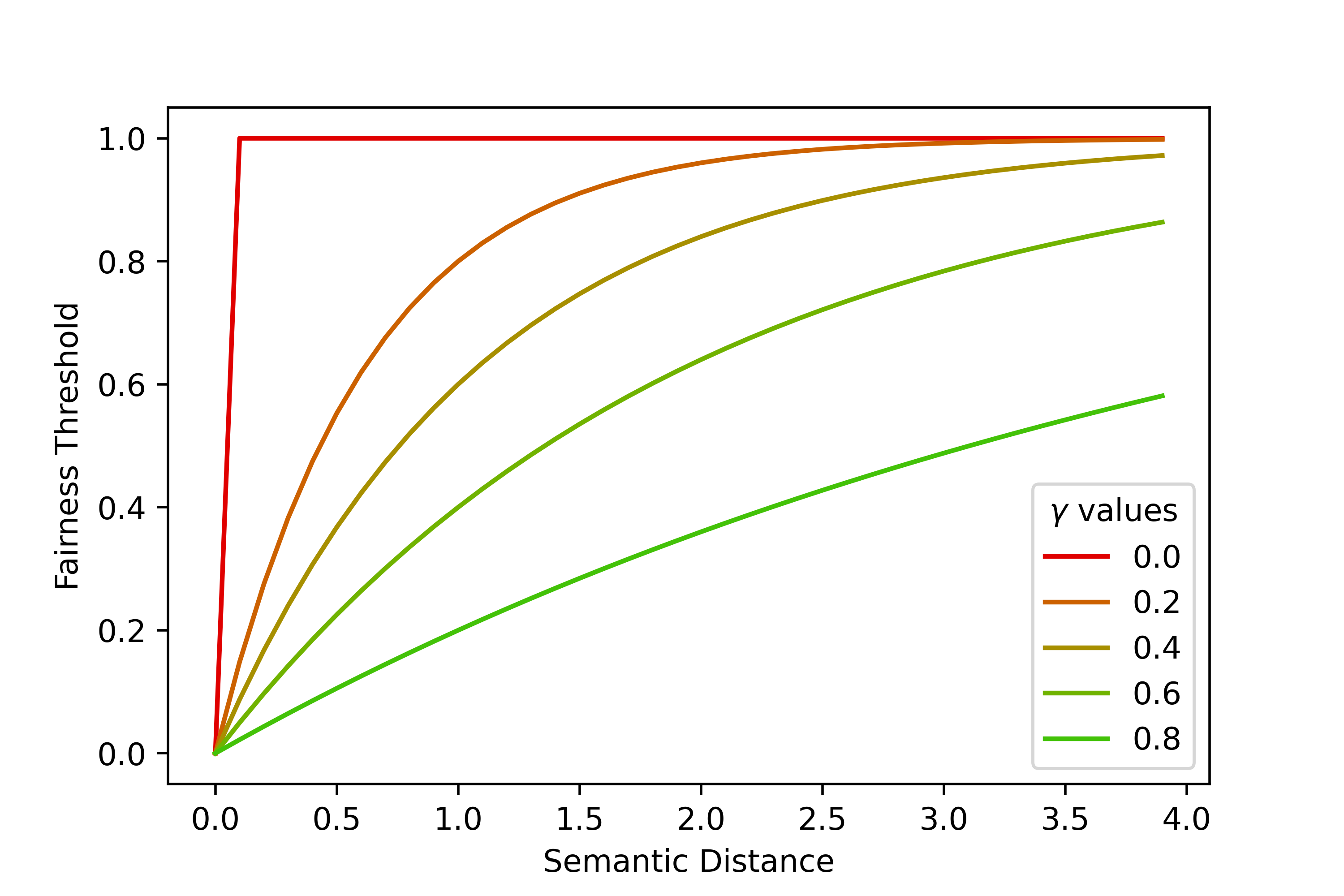}
\caption{Association-based fairness threshold curves on varying semantic distance with the identity elements for different $\gamma$ values}
\label{fig:dynFT}
\end{figure}

\begin{figure*}
    \centering
    \begin{subfigure}{0.32\textwidth}
         \centering
         \includegraphics[width=\textwidth]{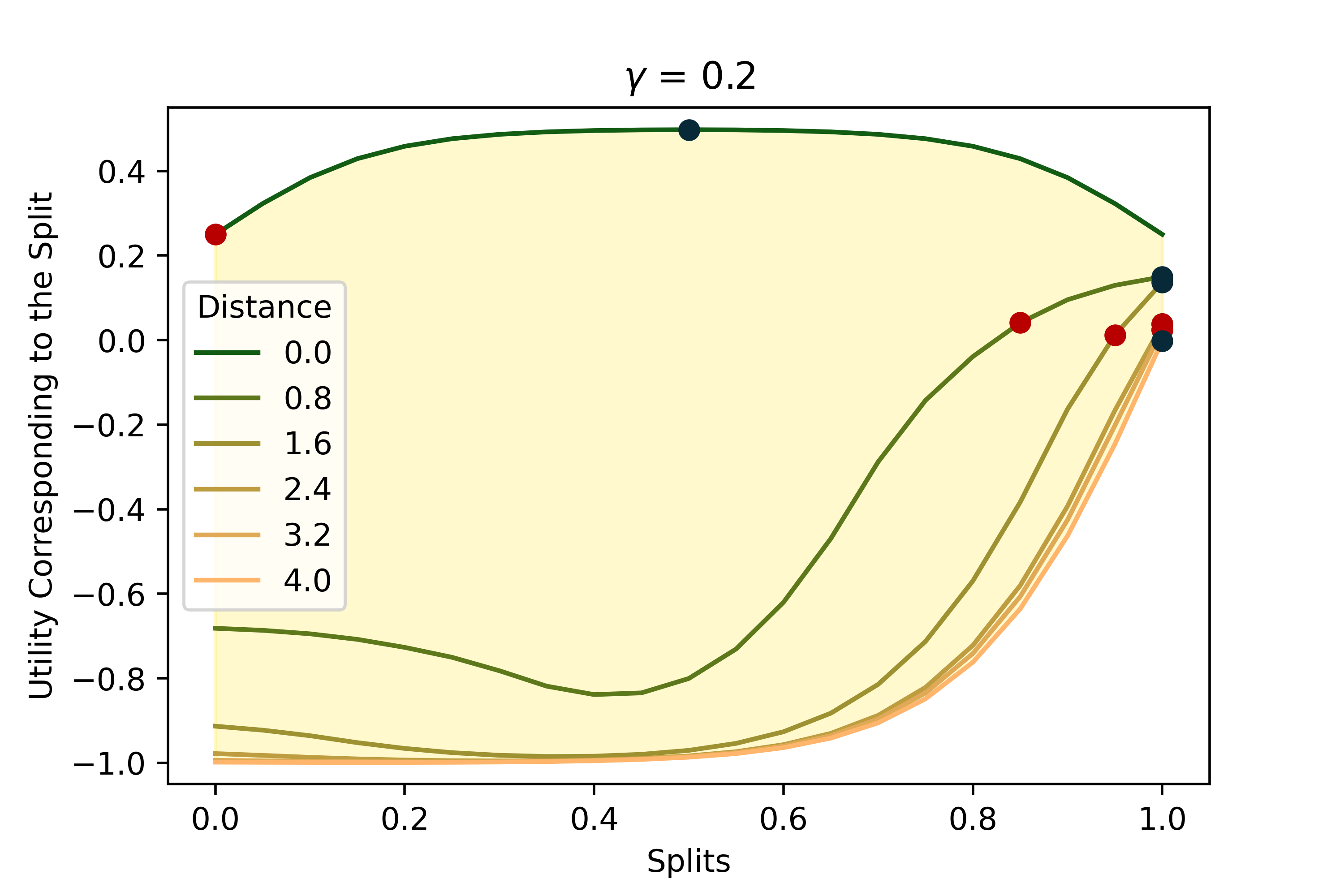}
    \caption{Low $\gamma$}
    \label{fig:utilmatStaticLowFT}
     \end{subfigure}
    \begin{subfigure}{0.32\textwidth}
         \centering
         \includegraphics[width=\textwidth]{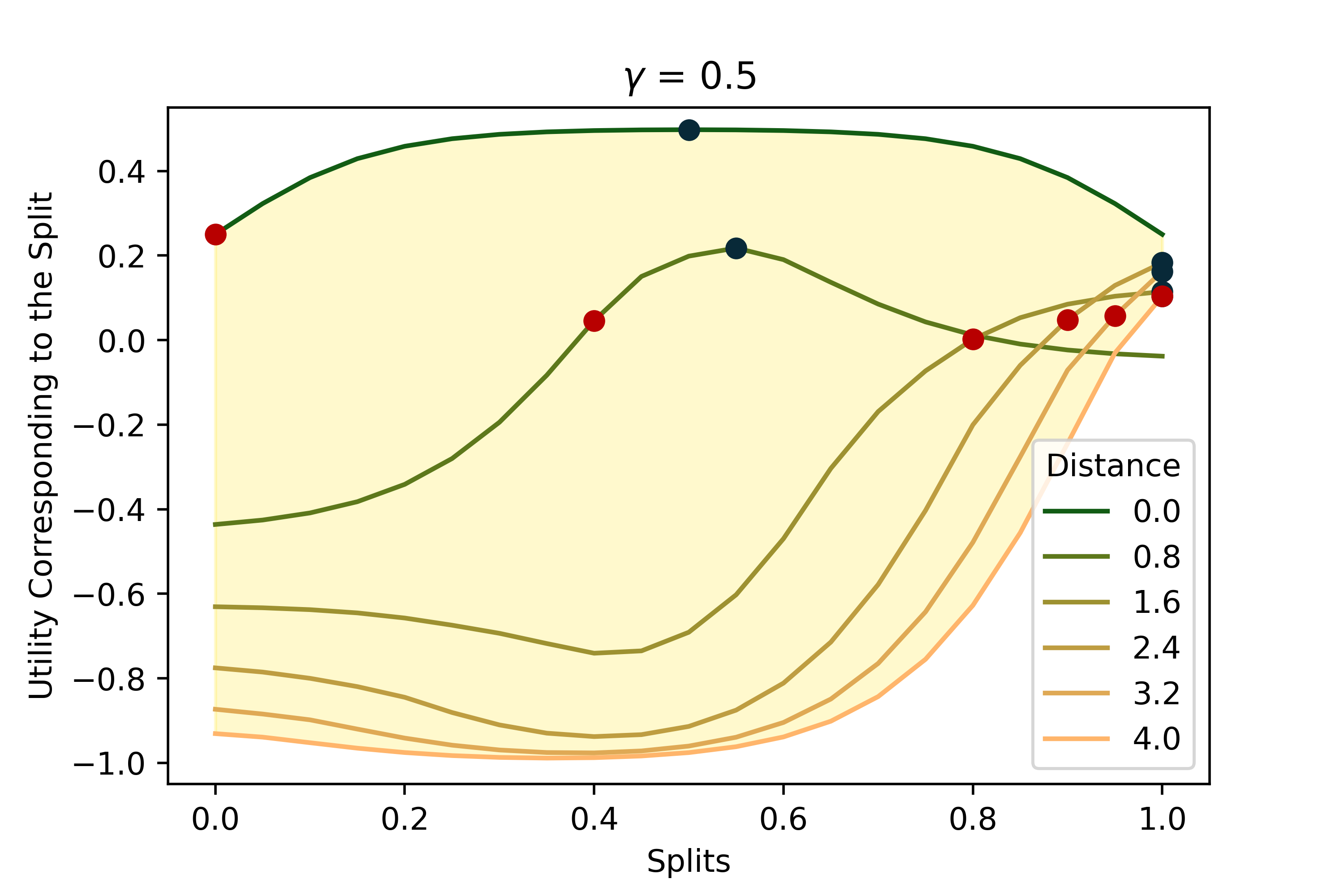}
    \caption{Moderate $\gamma$}
    \label{fig:utilmatStaticModFT}
     \end{subfigure}
     \begin{subfigure}{0.32\textwidth}
         \centering
         \includegraphics[width=\textwidth]{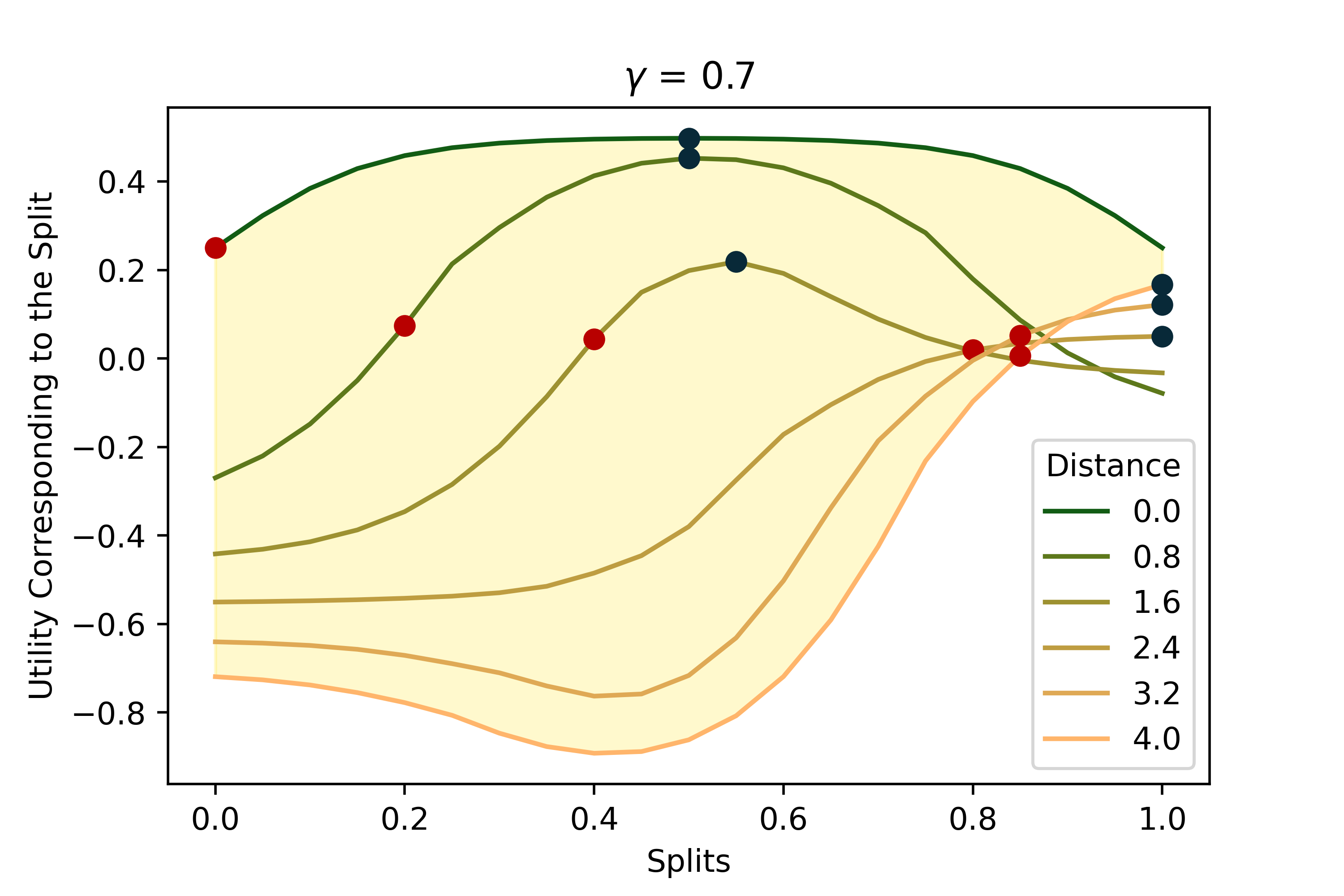}
    \caption{High $\gamma$}
    \label{fig:utilmatStaticHighFT}
     \end{subfigure}
    \caption{Utility plots for the agents with different values of $\gamma$}
    \label{fig:dynUtilCurve}
\end{figure*}

\begin{figure*}
    \centering
    \begin{subfigure}{0.32\textwidth}
         \centering
         \includegraphics[width=\textwidth]{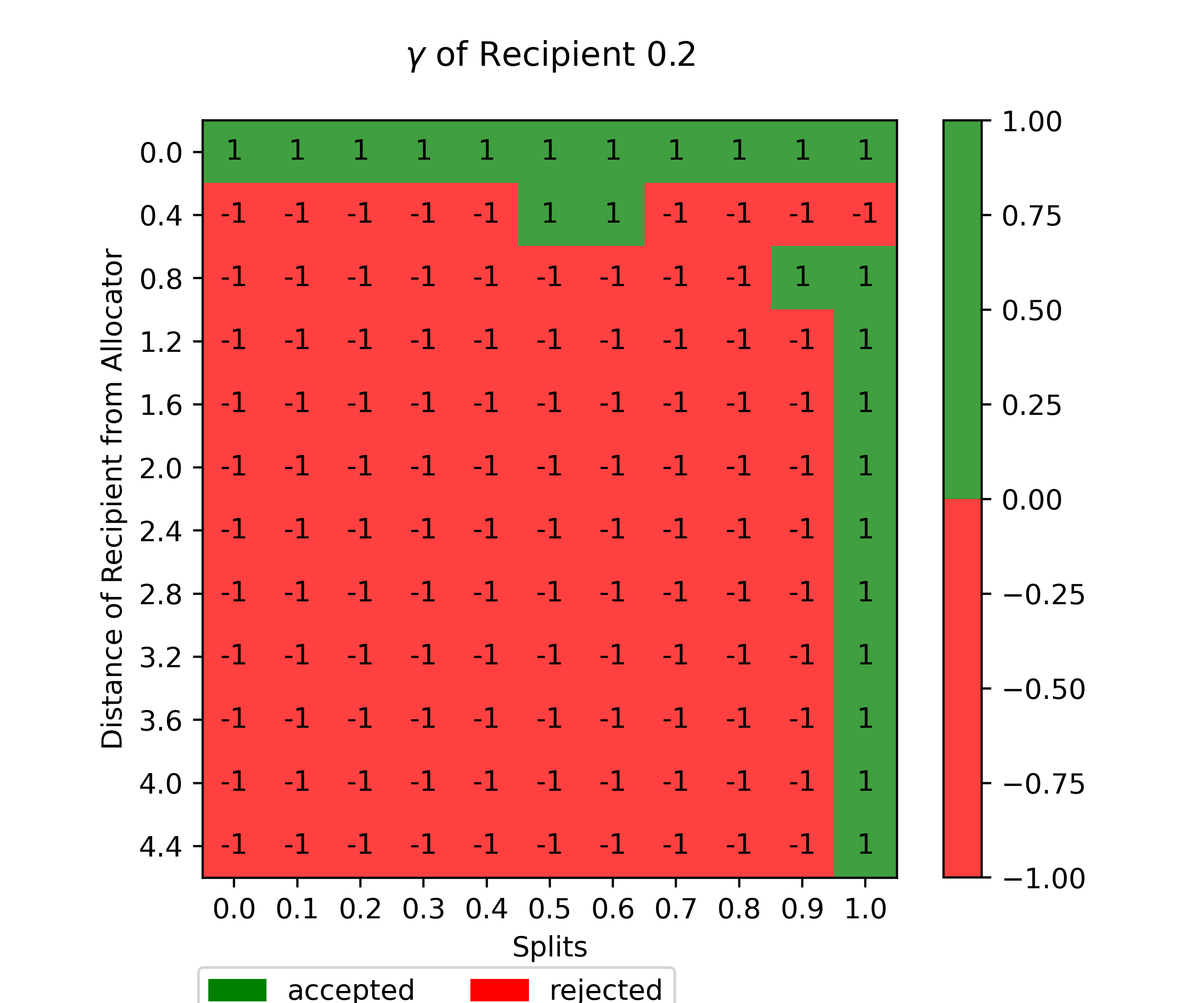}
    \caption{Low $\gamma$}
    \label{fig:matDynLG}
     \end{subfigure}
    \begin{subfigure}{0.32\textwidth}
         \centering
         \includegraphics[width=\textwidth]{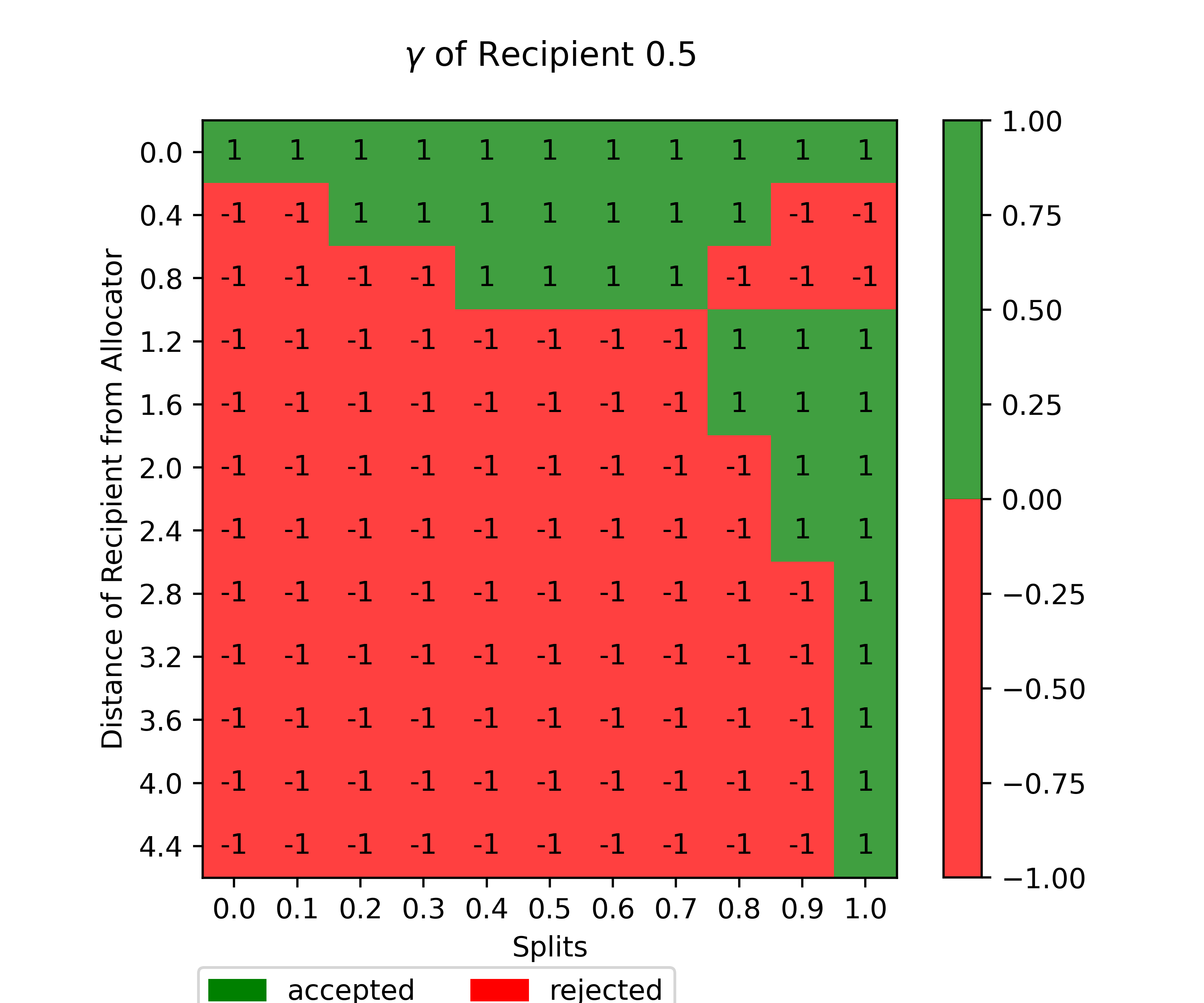}
    \caption{Moderate $\gamma$}
    \label{fig:matDynMG}
     \end{subfigure}
     \begin{subfigure}{0.32\textwidth}
         \centering
         \includegraphics[width=\textwidth]{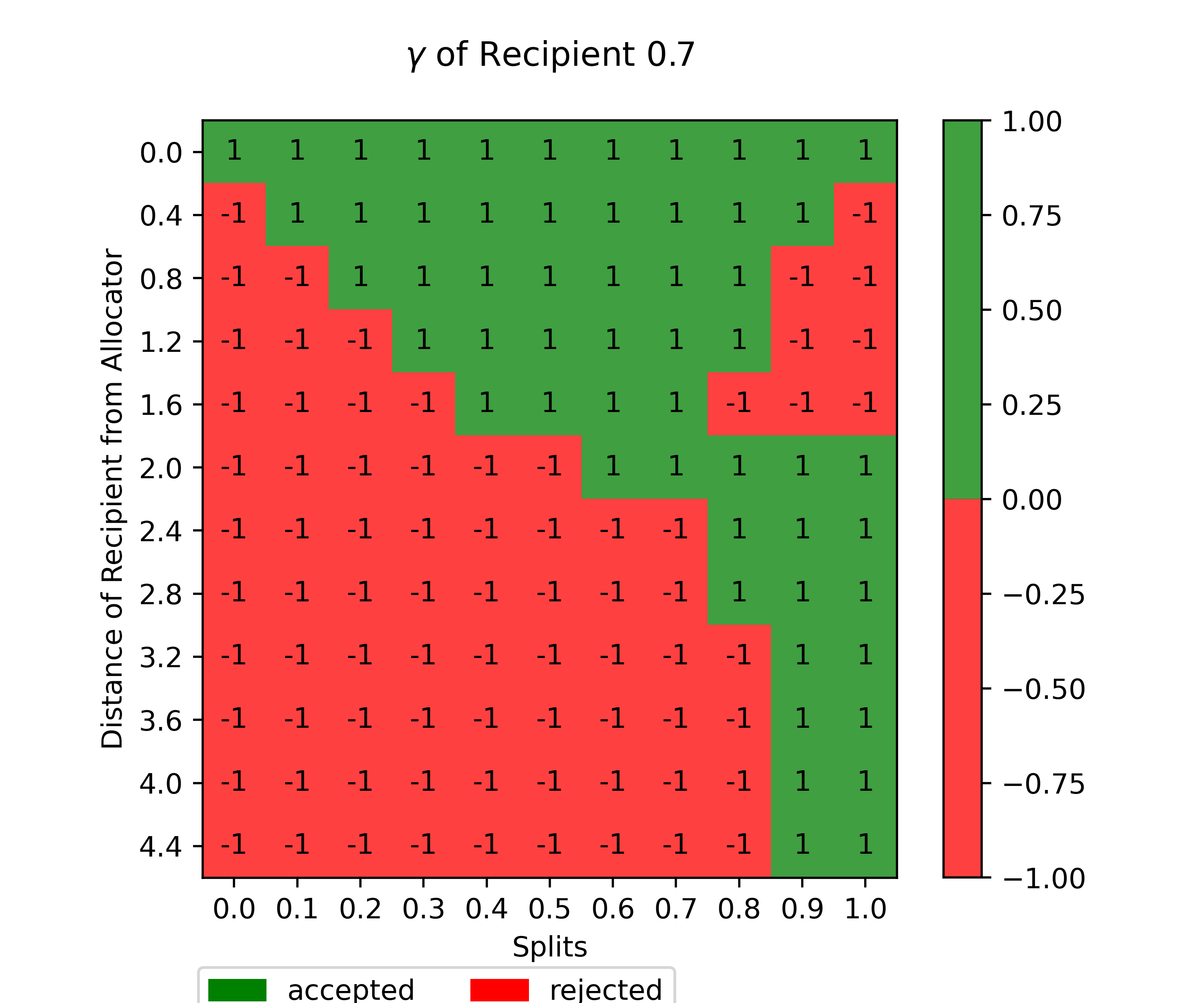}
    \caption{High $\gamma$}
    \label{fig:matDynHG}
     \end{subfigure}
     \caption{Acceptable Split Matrix with varying semantic distance of recipient with allocator for different values of $\gamma$}
    \label{fig:matRecDyn}
\end{figure*}

In this section, we introduce the notion of fairness using an association-based fairness threshold for the agents. It is similar to the semantic distance, d that the agents have with different aspects in their identity set. Agent a, has a fairness threshold $\tau$, for each element o in its identity set represented as $\tau_{a}(o)$. 

As discussed earlier, transcended agents account for the payoff of the aspects they identify with in their utility computation as described in Equation~\ref{eqn:ctutility}. Their fairness threshold should thus correlate with the extent of their transcendence level towards the other agent they are interacting with. The extent of transcendence of an agent with transcendence level $\gamma$ and semantic distance $d$ with the other agent is denoted using $\gamma^{d}$. Using this formulation, we define association-based fairness threshold ($\tau_{a}(o)$) towards aspect o in Equation~\ref{eq:associationFT} as follows:

\begin{equation}
    \tau_{a}(o) = 1 - \gamma_{a}^{d_{a}(o)}
    \label{eq:associationFT}
\end{equation}

Equation~\ref{eq:associationFT} can be interpreted as-- \textit{"The more you identify with an entity in your identity set, the less is your fairness threshold towards it"}. Formally, there is a negative correlation between the fairness threshold and the extent of identification. Next, using this equation, we plot the association-based fairness threshold in Figure~\ref{fig:dynFT} for different values of transcendence level $\gamma$. We observe that as an agent's semantic distance with an identity element increases, the value of $\tau$ also increases. For different values of $\gamma$, we observed that $\gamma$ contributes to the rate of change of $\tau$ with $d$. For increasing semantic distance $d$, $\tau$ of an agent with a low $\gamma$ value steadily increases, whereas $\tau$ of an agent with high $\gamma$ gradually increases. In this case, the utility of the agents is more significantly influenced by $\gamma$ and $d$. 

\subsection{Utility Plots}

We vary $\gamma$ and $d$ (and in turn $\tau$ which depends on these), to understand its effect on the utility of the agent. Figure~\ref{fig:dynUtilCurve} shows the utility plots for the agents for different values of $\gamma$. In this case, the nature of utility curves is more diverse as compared to Section~\ref{sec:staticFT} for agents with agent-based fairness threshold, because in this case, $\tau$ is also adapting with the change in the value of $d$. This also contributes to a larger yellow-shaded region in these utility plots denoting greater variation in utility. 

Like earlier, blue circles indicate the split at which the allocator gets the maximum utility and the red circles indicate the split at which the recipient gets its minimum acceptable utility. We observe that with the increase in $\gamma$, it takes a higher $d$ for the allocator to take the whole resource for itself. Also, as the recipient, with the increase in $\gamma$, the split that gives the minimum acceptable utility increases slowly as the distance increases.

\subsection{Acceptable Splits for the Recipient}

To understand more about the nature of deals that get accepted, we plot the matrices by varying the distance of the recipient from the allocator on the y-axis and the splits offered to it on the x-axis. The colour of cells in matrices indicates whether the deal is accepted as green or rejected as red. Figure~\ref{fig:matRecDyn} shows the matrices for different values of $\gamma$. We observe that at a lower distance, the recipient accepts almost all splits.
However, as $\gamma$ increases, the agents are more tolerant and accept splits greater than 0.5 up to low semantic distances to the allocator. 
 
We also note in Figure~\ref{fig:matDynLG} that the switch of the recipient to accept splits greater than 0.5 happens rapidly when the $\gamma$ of the agent is low, but it takes a higher distance for the recipient with high $\gamma$ to switch to selfish splits (Figure~\ref{fig:matDynHG}).

\section{Conclusions}

Allocation games are important to model scenarios of limited resource allocation. We noted that just incorporating responsible behaviour in agents is not sufficient for fair deals in this scenario. The baseline model of transcendence was found to be insufficient and unfair in the case of the allocation games since the agents factored in the utility of all the aspects they identified with, without considering the variation in the absolute value of utilities. The nature of allocation games characteristically differs from other game-theoretic scenarios like prisoners' dilemma, since it involves the distribution of a fixed resource, which can be split in multiple ways and yet the collective payoff of all the players always remains the same in all possible game states. Thus the notion of fairness is crucial to be modelled in agents which operate in scenarios representing the allocation games.

In this work, we focused on the Ultimatum Game (UG), which has been widely studied to better understand human behaviour in allocation scenarios. Though rational choice theories indicate that agents behave selfishly, empirical evidence suggests the contrary. We extended the identity-based model of agents - computational transcendence, to allocation games, and attempted to understand different factors which affect the resultant game states. We explored the interplay of the notion of fairness with subjective identity in the context of allocation games. Also, this model can be used to simulate agents with diverse behaviours and preferences similar to the variations observed in people across different cultures in the context of allocation games.

\section*{Acknowledgement}
We would like to thank the Machine Intelligence and Robotics (MINRO) Center funded by the Government of Karnataka, India and the Center for Internet of Ethical Things (CIET) funded by the Government of Karnataka, India and the World Economic Forum for funding and supporting this work.


\bibliography{main.bib}

\end{document}